\documentclass[usegraphicx,usenatbib]{mn2e}
\usepackage{amsmath}
\usepackage{amssymb}
\usepackage{times}
\usepackage{subfigure}
\usepackage{epsfig}
\usepackage{txfonts}

\usepackage{amsmath}
\usepackage{epsf}

\newcommand\rh{{r_{\rm h}}}
\newcommand\vhel{{v_{\rm HEL}}}
\newcommand\vgsr{{v_{\rm GSR}}}

\newcommand\objname{Segue 2 }
\newcommand\nameobj{\objname}

\newcommand\mg{$\Sigma$Mg }

\def\spose#1{\hbox to 0pt{#1\hss}}
\def\lta{\mathrel{\spose{\lower 3pt\hbox{$\sim$}}
    \raise 2.0pt\hbox{$<$}}}
\def\gta{\mathrel{\spose{\lower 3pt\hbox{$\sim$}}
    \raise 2.0pt\hbox{$>$}}}

\begin{document} 
\voffset-.6in

\title[Satellites of Satellites] {Segue 2: A Prototype of the
  Population of Satellites of Satellites}

\author[Belokurov et al]
 {V. Belokurov$^1$, M.G. Walker$^1$, N.W. Evans$^1$, G. Gilmore$^1$,
    M.J. Irwin$^1$, M. Mateo$^2$, \newauthor L. Mayer$^{3}$, E. Olszewski$^4$, 
    J. Bechtold$^4$, T. Pickering$^5$.
 \medskip
 \\$^1$Institute of Astronomy, University of Cambridge, Madingley
 Road, Cambridge, CB3 0HA, UK
 \\$^2$Department of Astronomy, University of Michigan, 
Ann Arbor, MI 48109, USA
 \\$^3$Institute for Theoretical Physics, University of Z\"urich, CH-8057
Z\"urich, Switzerland
 \\$^4$Steward Observatory, University of Arizona, Tucson,
  AZ 85721, USA
 \\$^5$MMT Observatory, University of Arizona, Tucson, USA
}


\maketitle

\begin{abstract}
  We announce the discovery of a new Milky Way satellite \nameobj
  found in the data of Sloan Extension for Galactic Understanding and
  Exploration (SEGUE). We followed this up with deeper imaging and
  spectroscopy on the Multiple Mirror Telescope (MMT). From this, we
  derive a luminosity of $M_v = -2.5$, a half-light radius of $34$ pc
  and a systemic velocity of $\sim -40$ kms$^{-1}$. Our data also
  provides evidence for a stream around \nameobj at a similar
  heliocentric velocity, and the SEGUE data show that it is also
  present in neighbouring fields. We resolve the velocity dispersion
  of \nameobj as 3.4 kms$^{-1}$ and the possible stream as $\sim 7$
  kms$^{-1}$. This object shows points of comparison with other recent
  discoveries, Segue 1, Boo II and Coma. We speculate that all four
  objects may be representatives of a population of satellites of
  satellites -- survivors of accretion events that destroyed their
  larger but less dense parents. They are likely to have formed at
  redshifts $z > 10$ and are good candidates for fossils of the
  reionization epoch.
\end{abstract}


\section{Introduction}

The idea that the outer parts of Galactic haloes are built up from the
merging and accretion of satellites is now well-established.  The
building blocks that contributed most to the Galactic halo have been
broken down into streams of debris.  Reconstructing the history is
difficult as the progenitors have been disassembled and phase
mixed. In Cold Dark Matter cosmogonies, smaller haloes form earlier
and are denser~\citep{Na97}. So, the entourage of the accreted
progenitors, smaller satellites of the bigger satellites, may have
survived against tidal destruction~\citep[see e.g.,][]{Di08}.  

Amongst the recent discoveries of Milky Way satellites, there are
objects whose properties are unlike conventional globular clusters or
dwarf galaxies, such as Coma, Segue 1 and Boo II~\citep{Be07,Wal07}.
They have half-light radii of $\sim 30$-$70$ pc and luminosities below
$M_v = -3$. In fact, Coma, Segue 1 and Boo II lie projected on the
Sagittarius stream, and have velocities consistent with Stream
membership.  Irrespective of whether they are dwarf galaxies or
globular clusters, it seems reasonable to conclude that they once
belonged to the Sagittarius galaxy. Could they be the first examples
of a population of satellites of satellites?

In this Letter, we report the discovery of a further object analogous
to Coma, Segue 1 and Boo II. Using the recently available SEGUE
imaging, we have extended our ongoing survey of stellar overdensities
in the outer Milky Way halo. We followed up the new object -- called
\nameobj -- with deep imaging and high resolution spectroscopy and
present its properties and possible genealogy here.

\begin{figure*}
\begin{center}
\includegraphics[width=0.85\textwidth]{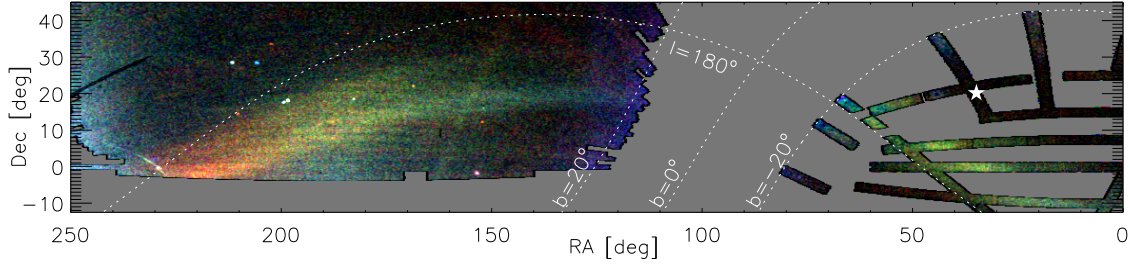}
\caption{\label{fig:asterisk} Location of \objname marked by asterisk
  with respect to ``the Field of Streams''.  in SEGUE imaging. This is
  a stellar density plot of all stars with $20.0<i <22.5$ and $-1< g-i
  <0.6$. The magnitude range is divided into three equal-sized bins
  analogous to \citet{Be06a}. Note that the Sagittarius Stream trailing
  arm is clearly visible crossing the equator at right ascensions
  $\alpha \sim 40^\circ$.}
\end{center}
\end{figure*}
\begin{figure*}
\begin{center}
\includegraphics[width=\textwidth]{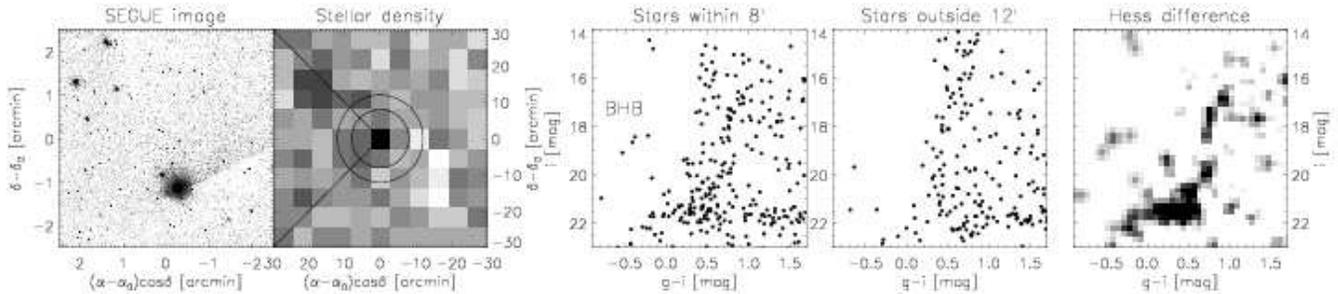}
\caption{Leftmost two panels: SEGUE image of the $5^\prime \times 5^\prime$
  field centered on \objname and density of all objects classified as stars
  in a $30^\prime \times 30^\prime$ field.  Note the bright saturated
  star and associated artefacts. The annuli are used in the
  construction of the CMDs. Rightmost three panels: CMD of the inner
  and outer regions, together with their Hess difference. There is a
  clear main sequence turn-off, together with sparse red giant branch
  (RGB) and blue horizontal branch (BHB). 
  \label{fig:aries_disc}}
\end{center}
\end{figure*}
\section{Data and Discovery}

The original Sloan Digital Sky Survey (SDSS) imaged most of the North
Galactic Cap plus three stripes of data in the South Galactic
Cap~\citep{Ab09}. The SEGUE survey~\citep{Ya09} is primarily
spectroscopic, but complements the SDSS imaging data with 15
$2.5^\circ$ wide stripes along constant Galactic longitude, spaced by
approximately $20^\circ$ around the sky.  The stripes probe the Galaxy
at a variety of longitudes, sampling the changing relative densities
of thin disk, thick disk, and halo.  The SEGUE imaging footprint is
illustrated at {\tt http://www.sdss.org/dr7/seguephoto{\_}big.gif}

Figure~\ref{fig:asterisk} shows the familiar picture of the ``Field of
Streams''~\citep{Be06a}, together with the SEGUE imaging stripes with
right ascension $\alpha < 90^\circ$ and declination $\delta <
50^\circ$. We have also cut on Galactic latitude $b < -20^\circ$ to
avoid showing regions dominated by the Galactic disk. As is usual, the
magnitudes of stars have been corrected for extinction using the maps
of \citet{Sc98}. Running the algorithm for identification of
overdensities described in \citet{Be06b}, we isolate high
significance peaks. The location of the highest is marked by a white
asterisk in Figure~\ref{fig:asterisk}.

Figure~\ref{fig:aries_disc} shows the SEGUE view of the overdensity.
The leftmost panel is a cut-out of the sky with the overdensity at the
centre. The only discernible structure is in fact an unrelated
saturated foreground star and accompanying artefacts. In the next
panel, the density of resolved stars is shown, and now there is an
evident overdensity with significance~\citep{Ko08} $S = 4.7$ at the
centre of the image. Annuli are marked which are used to select stars
within $8^\prime$ and outside $12^\prime$ for the two color magnitude
diagrams (CMDs). The rightmost panel gives the Hess difference, and
there is a clear main-sequence turn-off, together with definite hints
of a red giant branch (RGB) and a blue horizontal branch (BHB). 

\begin{figure}
\begin{center}
\includegraphics[width=0.4\textwidth]{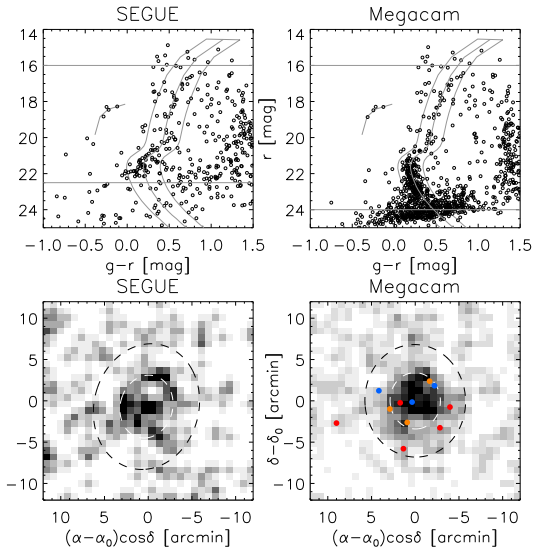}
\caption{Top left and right; CMDs of \nameobj in SEGUE and in the
  follow-up Megacam data, which goes at least 2 mag deeper. The
  follow-up data shows a well-defined main sequence, with grey lines
  marking the ridgeline of M92 and the mask used to select members.
  Bottom left and right: Grey-scale density of stars selected by in
  the $24^\prime \times 24^\prime$ field of view centered on
  \objname. The ellipses show the model isodensity contours
  corresponding to $\rh$ and $2\rh$. The colored dots show the
  locations of likely members followed up spectroscopically.}
\label{fig:aries_megacam}
\end{center}
\end{figure}

\begin{figure*}
\begin{center}
\includegraphics[width=0.8\textwidth]{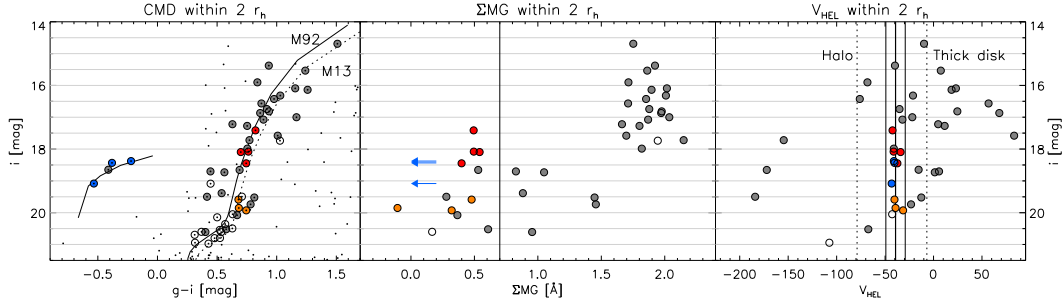}
\caption{Left: CMD of all stars (black dots) within 2$\rh$, with
  objects followed-up spectroscopically circled. Filled circles with
  valid velocities and \mg color-coded such that gray shows the
  foreground population, red shows red clump giants, blue show BHBs
  and orange subgaints in \objname. The M92 ridgeline (including the
  BHB) is overplotted together with the M13 ridgeline for
  comparison. Middle: The distribution of \mg values, showing clean
  separation between dwarfs in the thick disk (high \mg) and giants in
  \objname (low \mg) at bright magnitudes. The vertical solid line
  marks the boundary adopted here. Note that at fainter magnitudes,
  there is a population of stars with intermediate \mg and velocities
  similar to the systemic velocity of \objname. Right: Heliocentric
  radial velocities of all stars with good spectra. The characteristic
  velocities associated with the halo and thick disk at this location
  are enclosed by vertical dashed lines, together with solid lines
  showing the systemic velocity of \objname plus or minus 10
  kms$^{-1}$.}
\label{fig:aries_spec}
\end{center}
\end{figure*}

\section{Follow-Up}

\subsection{Imaging}

Follow-up imaging of \nameobj was carried out on Oct 7 2007 using the
Megacam imager (Mcleod et al 2006) on the Multiple Mirror Telescope
(MMT).  Megacam comprises 36 2048x4608 E2V CCDS. With 2x2 binning,
pixels are $0.16^{\prime\prime}$ and each image is $24^\prime \times
24^\prime$ in size.  Six 300s exposure images in $r'$ and 7 in $g'$
were collected with dithers of 100-200 pixels in each coordinate
between frames.  Frames were processed using SAO's MEGARED package and
combined using SWARP package in TERAPIX \citep{Ra01}.  A final set of
object catalogs was generated from the stacked images and objects were
morphologically classified as stellar or non-stellar (or noise-like).
The detected objects in each passband were then merged by positional
coincidence (within $1^{\prime\prime}$) to form a combined $g,r$
catalogue and photometrically calibrated on the SDSS system using
stars in common.

Figure~\ref{fig:aries_megacam} compares the original SEGUE data with
the Megacam follow-up. From the CMDs, it is immediately clear that the
Megacam data probe at least 2 magnitudes deeper and reveal the main
sequence of \nameobj. The algorithm of \citet{Ma08} is applied to
stars selected by the masks shown in the Figure. Model isodensity
contours, based on both the SEGUE and Megacam data, are shown in the
lower panels.  We see that the shallower SEGUE data yields a slightly
more extended and elliptical light profile.  The extracted structural
parameters using the deeper Megacam data are listed in
Table~\ref{tab:struct}.  There are possibly 4 BHB stars associated
with \nameobj, which can be used to obtain a distance modulus $m-M =
17.7$ or 35 kpc. Using this, the half-light radius is $\rh = 34$ pc
which is small compared to typical ultrafaint dwarf galaxies
~\citep{Be07, Gi07}.

\begin{table}
\centering
\begin{tabular}{lc}
\hline
Property & \null \\
\hline
Coordinates (J2000) & $\alpha =02:19:16$, $\delta = 20:10:31$ \\
Coordinates (Galactic) & $\ell = 149.4^\circ$, $b= -38.1^\circ$ \\
$v_\odot$, $\sigma$ & $-39.2 \pm 2.5$, $3.4^{+2.5}_{-1.2}$ kms$^{-1}$\\
 Position Angle & $182^{\circ} \pm 17^{\circ}$\\
 Ellipticity & $0.15 \pm 0.1$\\
$\rh$ (Exponential) & $3\farcm4 \pm 0\farcm2$\\
(m$-$M)$_0$ & $17\fm7 \pm 0\fm1$\\
M$_{\rm tot,V}$ & $-2\fm5 \pm 0\fm2$\\
\hline
\end{tabular}
\caption{Properties of the \objname Satellite \label{tab:struct}}
\end{table}

\subsection{Spectroscopy}

On 22-23 and 26 October 2008, we obtained high resolution spectra of
358 targets around \nameobj using three independent fiber
configurations with the Hectochelle spectrograph at the MMT.  Over the
$1^\circ$ field, we targeted RGB candidates as faint as $i\sim 21.5$,
as well as the handful of BHB candidates at $i\sim 18.5$.  We
extracted and calibrated spectra following the procedure described
by~\citet{Mat08}.  The Hectochelle spectra span $5150-5300$ \AA, where
RGBs and foreground dwarfs prominently exhibit the Mg-I/Mg-b triplet
(MgT) absorption feature.  For the RGB candidates, we measure velocity
by cross-correlating each spectrum against a high signal-to-noise
template spectrum, built from co-added spectra of late-type radial
velocity standards observed with Hectochelle.  However, this template
poorly resembles the spectrum of a BHB, in which high temperature
suppresses Mg absorption and the only prominent absorption feature is
the FeI/FeII blend at 5169 \AA.  Thus, for each BHB candidate, we
measure the centroid of this feature and calculate velocity directly
from the redshift.  For all spectra, we also measure a composite
magnesium index, $\Sigma$Mg, effectively a pseudo-equivalent width for
the MgT~\citep{Wa07}.  This quantity correlates with metallicity,
temperature and surface gravity, and helps to separate members from
foreground.  We determine measurement errors for both velocity and
$\Sigma$Mg using the bootstrap method described by~\citet{Wa09}.
Velocities and \mg in the \nameobj sample have mean (median) errors of
1.1 (0.6) km s$^{-1}$ and 0.20 (0.17) \AA\ , respectively.

Figure~\ref{fig:aries_spec} shows correlations between the photometric
and spectroscopic properties of stars within $2\rh$. The first panel
shows the locations of stars targeted for follow-up on the CMD,
together with the ridgelines of M92 ([Fe/H] = -2.24) and M13 ([Fe/H] =
-1.65) from \citet{Cl08}. The second and third panels show \mg and
$\vhel$ plotted against $i$ band magnitude, from which we will identify
three classes of \nameobj members, namely bright red clump giants
(RCGs), fainter RGBs and BHBs, and two types of contaminants, Galactic
foreground and possible tidal stream.

\begin{figure}
\begin{center}
\includegraphics[width=0.4\textwidth]{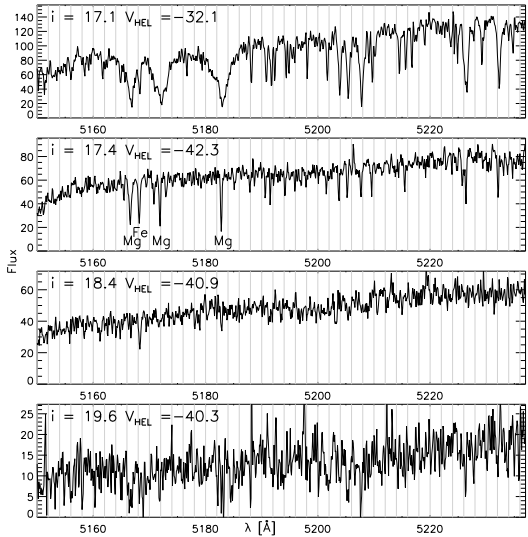}
\caption{Examples of spectra of thick disk contaminant (top), together
  with three members of \nameobj, a RCG (upper middle), a BHB (lower
  middle) and a faint RGB (bottom). Only the bluer part of the
  wavelength range containing the MgT is shown. }
\label{fig:aries_spec_eg}
\end{center}
\end{figure}

One population clearly stands out in the \mg panel. Given the
magnitude distributions, it is evident that the stars with higher \mg
are dwarfs in the Galactic thick disk. At a similar magnitude $i \sim
17.5$, there is only one other population with a tight \mg
distribution, namely the red clump stars visible in the first
panel. These form a narrow distribution in velocity space, suggesting
that the systemic velocity is $\sim -40$ kms$^{-1}$.  Sample spectra
of a Galactic dwarf and a RCG are shown in
Figure~\ref{fig:aries_spec_eg}. Note that the dwarf has higher surface
gravity and potentially higher metallicity than the RCGs, and hence
has broader absorption features.  The velocity signature is
corroborated by the three BHBs, which all have velocities $\sim -40$
kms$^{-1}$. Finally, we use the systemic velocity of \nameobj with a
range of $\pm 10$ kms$^{-1}$ as a secondary cut, as shown by the
vertical lines in the third panel. This gives a further three
candidate members (shown in yellow), which all lie redwards of the M92
ridgeline and are better described by more metal-rich templates like
M13. An example of their spectra is illustrated in the fourth panel of
Figure~\ref{fig:aries_spec_eg}. Unfortunately, the low signal-to-noise
does not allow the direct extraction of the metallicity for these
stars.

\begin{figure}
\begin{center}
\includegraphics[width=0.5\textwidth]{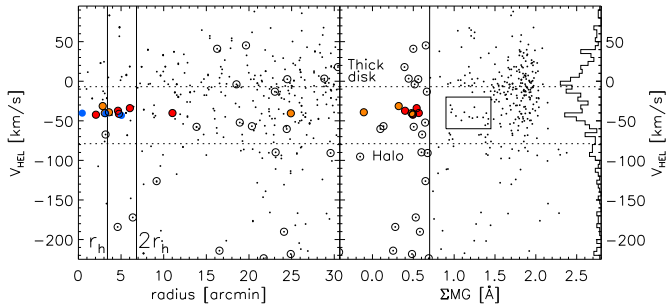}
\caption{Heliocentric radial velocity versus elliptical radius (left)
  and \mg (right). All the stars with measured \mg are shown as black
  dots, whilst stars satisfying \mg $< 0.7$ are circled. The extent of
  \objname is illustrated by the vertical lines showing $\rh$ and
  $2\rh$. The open circled stars do not appear to be kinematically
  associated with \objname. The structural parameters inferred from
  the photometry are consistent with the spectroscopic and kinematic
  signal. Note that the velocity histogram on the extreme right
  reveals a substantial population of stars with $-60 <\vhel<-30$
  kms$^{-1}$ that cannot be attributed to either thick disk or halo.}
\label{fig:aries_dist}
\end{center}
\end{figure}
\begin{figure}
\begin{center}
\includegraphics[width=0.4\textwidth]{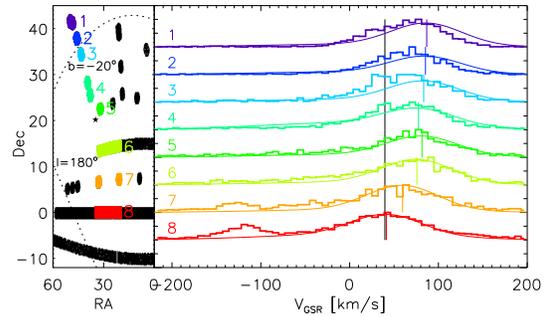}
\caption{\label{fig:aries_stream} Left: SEGUE spectroscopic fooprint in
  the vicinity of \objname. Right: Histograms of radial velocities
  corrected to the Galactic Standard of Rest in fields coded according
  to color. The black vertical lines shows the systemic velocity of
  \objname, whilst the colored ones show the mean velocity of the
  thick disk. The smooth curves show the sums of Gaussian velocity
  distributions of thick disk ($\langle v \rangle =180, \sigma = 50$
  kms$^{-1}$) and halo ($\langle v \rangle =0, \sigma = 100$
  kms$^{-1}$). Note the prominent bumps in the red and orange
  histograms corresponding to the Sagittarius trailing arm, and
  similarly the bumps in the pale blue and green histograms.}
\end{center}
\end{figure}

These three fainter giant stars may be representatives of a distinct
population with low to intermediate \mg, clearly visible in the second
panel.  The velocities of the stars in this population are offset from
the thick disk, but roughly centered on \objname. The population is
much more evident in Figure~\ref{fig:aries_dist}, which shows all
stars with measured velocities and \mg within $45'$. The velocity
histogram along the rightmost vertical of the plot shows that the
foreground consists of the expected thick disk and halo populations,
together with an unknown component, perhaps a tidal stream. The stars
in this component extend over the entire field and they are
kinematically colder than the thick disk but significantly hotter than
\nameobj.

The left panel of Figure~\ref{fig:aries_stream} shows the SEGUE
spectroscopic footprint around \nameobj. The right panel shows the
distributions of radial velocities corrected to the Galactic Standard
of Rest are shown for each color-coded field. The underlying smooth
curves come from a simple model of the Galactic thick disk and halo
represented as two Gaussians with means and dispersions of ($\langle v
\rangle =180, \sigma = 50$ kms$^{-1}$) and ($\langle v \rangle =0,
\sigma = 100$ kms$^{-1}$) respectively. We immediately notice the
presence of bumps or cold features in several fields. At low
declination, the Sagittarius trailing arm is detected at $\vgsr
\approx -120$ kms$^{-1}$. At higher declinations, there are
significant deviations from the smooth curves at velocities $\vgsr
\approx 40$ kms$^{-1}$, which is close to the systemic velocity of
\nameobj. 

More speculatively, we suggest that the stream-like overdensity seen
in the MMT data is part of the same structure as the features seen in
the velocity histograms in the higher declination (lower Galactic
latitude) SEGUE fields. This could be confirmed with distance
estimates to the structures. There are many possibilities as to the
nature of this overdensity. First, it could be part of the Sagittarius
Stream, which lies in the same area of the Sky. However, the bulk of
the Sagittarius debris lies at a lower declination (see
Figure~\ref{fig:asterisk}). Another possibility is the Monoceros
ring~\citep{Ya03}, visible in the SDSS data at the same Galactic
longitude but positive latitude. However, the kinematic feature in the
SEGUE fields seems too localized and is limited to latitudes
$-30^\circ < b < -20^\circ$. Finally, \citep{Ma04} reported the
detection of an extended structure -- perhaps a segment of tidal
debris -- a few degrees away from these fields as part of their survey
of the Andromeda galaxy. Irrespective of which possibility is correct,
our hypothesis is that \nameobj is embedded in a tidal stream.

\section{Physical Properties and Stellar Population}

Assuming \nameobj has a Gaussian velocity distribution, we measure its
velocity dispersion using a maximum-likelihood method.  Specifically,
we evaluate the marginal likelihood obtained after integrating the
usual Gaussian likelihood over all mean velocities (see Equations 1
and 2 of~\citet{Kl04}). We find the velocity dispersion that maximizes
this likelihood and determine the boundaries that enclose 68\% and
95\% of the area under the likelihood curve.  For the five bright RCGs
(marked in red in Fig.~\ref{fig:aries_dist}), we measure a velocity
dispersion $\sigma_{V_0}=3.4_{-1.2(-2.2)}^{+2.5(+8.2)}$ km s$^{-1}$.
This result is not strongly sensitive to our membership criteria -- if
we include the three fainter candidates (orange points in the Figures)
passing our initial velocity and magnesium cuts we obtain
$\sigma_{V_0}=3.6_{-1.0(-2.3)}^{+1.7(+4.5)}$ km s$^{-1}$.

Our analysis indicates that we resolve the central velocity dispersion
of \nameobj, ruling out zero with more than 99\% confidence.  Adopting
the idealised assumptions of spherical symmetry, dynamical equilibrium
and ``mass follows light,'' implicit in the formula $M= 850
r_{h}\sigma_{V_0}^2$~\citep{Il76, Si07}, the velocity dispersion of
the five bright RCG members implies a dynamical mass of
$M=5.5_{-3.1(-4.3)}^{+10.9(+52)}\times 10^5 M_{\odot}$, and a
mass-to-light ratio of $M/L_V=650_{-380(-520)}^{+1300(+6200)}
[M/L_V]_{\odot}$.

We can also estimate the kinematic properties of the stream from 15
prospective stream members satisfying $-60 \leq V\leq -20$ km s$^{-1}$
and $0.9 \leq \Sigma\mathrm{Mg}\leq 1.45$ shown as a box in
Figure~\ref{fig:aries_dist}. The mean velocity of the stream is $-45.1
\pm 0.1 (\pm 0.2)$ kms$^{-1}$ with 1 $\sigma$ (2 $\sigma$)
errors. This is obtained by marginalizing over the dispersion. It is
offset by $\sim 5$ kms$^{-1}$ from the systemic velocity of \nameobj.
The velocity dispersion of the stream is $7.1_{-1.2(-2.1)}^{+1.8
  (+4.2)}$ kms$^{-1}$.

For the brightest members of \nameobj (the BHB and RCG stars), the
signal-to-noise of the spectra is good enough to estimate
metallicity. This was done by directly comparing the average continuum
normalised spectrum of the three BHB spectra satisfying $18.0 < g <
19.0$ and the average of the five best RCG spectra satisfying $18.0 <
g < 19.2$, with a grid of model atmosphere spectra (see \citet{Wa09}
for further details).  Using the relationship $\log_{10}(T_{\rm eff})
= 3.877 - 0.26(g-r)$ \citep{Iv06}, the average color ($\langle
g-r\rangle = 0.53$) of the five RCG members implies $\langle T_{\rm
  eff}\rangle = 5500$K for these stars, while for the BHB stars the
average color ($\langle g-r\rangle = -0.22$) implies $\langle T_{\rm
  eff}\rangle = 8600$K. Anchoring the $T_{\rm eff}$ for the spectral
fit tightens constraints on gravity and metallicity, giving $\log g
=2.5 \pm 0.5$ and [Fe/H] $= -2.0 \pm 0.25$ for the averaged BHB stars
and $\log g =2.5 \pm 0.5$ and [Fe/H] $= -2.0 \pm 0.25$ for the
averaged RCG stars. For the latter, each had sufficiently strong Mg
and Fe lines and continuum signal-to-noise to model individually.  In
all cases the best fit model spectrum satisified $2.0 \le \log g \le
3.0$; $5000\rm{K} \le T_{\rm eff} \le 5500$K and $-2.5 \le $[Fe/H]$
\le -2.0$ corroborating their categorisation as RCG stars.  These
colors, magnitudes, surface gravities and effective temperatures are
all fully consistent with the BHB and RCG stars being at a common
distance (see, e.g., \citealt{Gr05}, p. 57).

\section{Conclusions}

A search in the high Galactic latitude ($|b| > 20^\circ$) area covered
by SEGUE imaging has revealed another new satellite.  \nameobj has a
half-light radius of $\sim 30$ pc and an absolute magnitude of $M_v =
-2.5$. The photometry and spectroscopy suggest that the metallicity
[Fe/H] $\sim -2$. \nameobj is similar in structure, size, luminosity
and velocity dispersion to three other recent discoveries, namely
Segue 1, Boo II and Coma.

The latter three are likely to be embedded in the Sagittarius Stream.
For example, Coma is superposed on the edge of the Stream and is at a
distance that suggests association with the old leading (C) arm of the
Sagittarius. \citet{Ni09} shows that Segue 1 stars are
indistinguishable from the Sagittarius Stream, both photometrically
and kinematically. It is natural to conclude that Segue 1 was once a
satellite of the Sagittarius galaxy. \citet{Ko09} have shown that Boo
II lies close to the young leading (A) and old trailing (B) arms of
the Sagittarius, and has similar kinematics. 

There is indirect evidence that our new discovery \nameobj is also
immersed in a stream. To begin with, it lies on the edge of the
Sagittarius stream, as seen by the SEGUE survey. Kinematically, there
is a cold stream-like component in both our follow-up spectroscopy and
in the SEGUE spectroscopy of nearby fields.

We speculate that all four objects are possibly satellites of
satellites, remnants from the disruption of larger galaxies in the
Milky Way halo.  Under the simple assumption of an isothermal halo,
the current circular velocity of \nameobj is $v_{\rm circ} \sim 5$
kms$^{-1}$. The original $v_{\rm circ}$ before the tiny dwarf accreted
onto the Milky Way halo as part of a parent subgroup might have been
close to $10-15$ kms$^{-1}$ (a reducton by a factor of $2-3$ is
expected as a result of tidal shocks -- see Mayer et
al. 2007). Initial halo masses corresponding to such circular
velocities are below $10^7 M_{\odot}$.  With such low masses, these
satellites of satellites are likely to have formed before
reionization, at $z > 10$, since afterwards gas would have been
photoevaporated owing to heating by the cosmic ionizing background
(Barkana \& Loeb 1999). Such an early formation epoch would naturally
imply a high central density and a resulting resilience to tidal
disruption of their inner core. The inner, surviving core of the
object is what we may be witnessing in these four objects.  These, and
not the more luminous, ordinary dwarf spheroidal satellites of the
Milky Way and M31, are good candidates for being fossils of the
reionization epoch (Diemand et al. 2007).  Accounting properly for
their origin in the context of the formation of the Local Group is
crucial in order to interpret correctly the discrepancy between the
observed luminosity function of satellites and the predicted
substructure mass function. Satellites of satellites are only resolved
in the most recent dark matter cosmological simulations of the Milky
Way halo (Diemand et al. 2008), but not yet in more realistic fully
hydrodynamical simulations that account for the various mechanisms to
which baryons are subject. Being able to identify satellites of
satellites in cosmological hydrodynamical simulations will thus be a
crucial step for interpreting new observations such as those presented
here.

\section*{Acknowledgements}
Funding for the SDSS and SDSS-II has been provided by the Alfred P.
Sloan Foundation, the Participating Institutions, the National Science
Foundation, the U.S. Department of Energy, the National Aeronautics
and Space Administration, the Japanese Monbukagakusho, the Max Planck
Society, and the Higher Education Funding Council for England. The
SDSS Web Site is http://www.sdss.org/.  Some of the observations
reported here were obtained at the MMT Observatory, a joint facility
of the Smithsonian Institution and the University of Arizona.
          
This work is based (in part) on data products produced at the TERAPIX
data center located at the Institut d'Astrophysique de Paris.  MM
acknowledges support from NSF grants AST-0206081 0507453, and
0808043. EO acknowledges support from NSF Grants AST-0205790, 0505711,
and 0807498.

\end{document}